\documentclass{vhepu14}

\def\Journal#1#2#3#4{{#1} {\bf #2}, #3 (#4)}

\def\AA{\em Astronomy \& Astrophysics}
\def\APP{\em Astroparticle Physics}
\def\Scie{\em Science}
\def\ApJ{\em Astrophysical Journal}
\def\EA{\em Exp. Astronomy}

\def\JCAP{\em Journal of Cosmology and Astroparticle Physics}
\def\MN{\em M.N.R.A.S}
\def\Nat{\em Nature}

\def\PRL{\em Phys. Rev. Lett.}

\def\SSR{\em Space Sci. Rev.}

\def\be{\begin{equation}}
\def\ee{\end{equation}}
\def\bea{\begin{eqnarray}}
\def\eea{\end{eqnarray}}

\newcommand  \ga {${\gamma}$}
\newcommand  \gag	{$\gamma-\gamma$}

\newcommand  \mic {$\mu$m}

\newcommand   \gray {$\gamma-$ray}

\newcommand    \fermilat  {\textit{Fermi}-LAT}
\newcommand   \veritas   {VERITAS}
\newcommand   \magic   {MAGIC}
\newcommand   \hess   {H.E.S.S.}

\newcommand{\Photo}{\includegraphics[height=35mm]{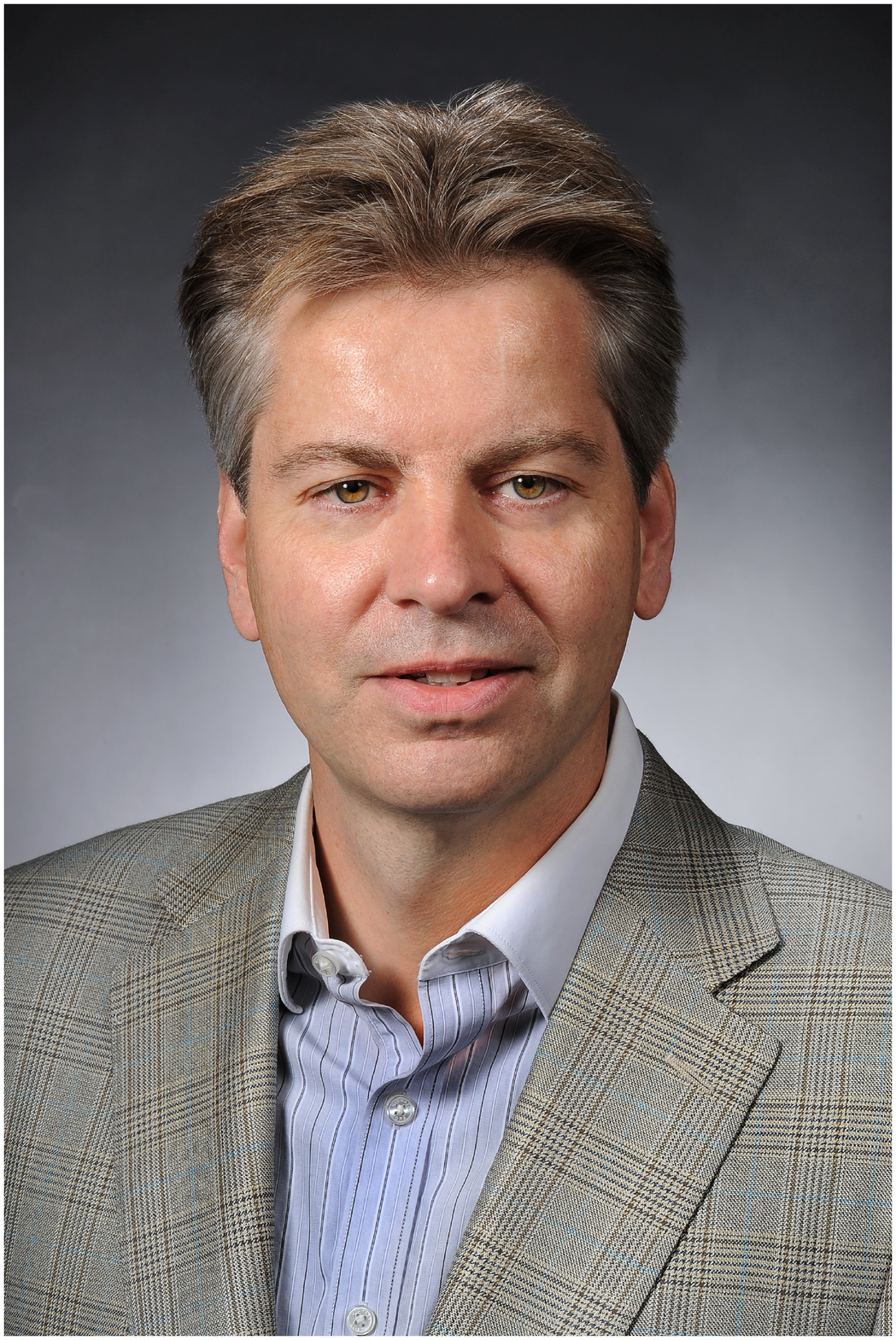}}

\begin{document}
\vspace*{4cm}
\title{THE GAMMA RAY OPACITY OF THE UNIVERSE -- INDIRECT MEASUREMENTS OF THE EXTRAGALACTIC BACKGROUND LIGHT}
\author{ F. KRENNRICH}
\address{Department of Physics and Astronomy, Iowa State University, \\ Zaffarano Hall, Ames, IA 50011-3160}

\maketitle\abstracts{
Indirect constraints on the intensity of the Extragalactic Background Light (EBL) were provided by  recent studies of  extragalactic sources emitting sub-TeV to multi-TeV photons.  These constraints are provided thanks to the absorption of  \ga\ rays by soft photons from the EBL (UV/optical/IR) via $\rm   e^{\pm} $ pair production by  \gag\ interactions.   This paper provides an overview of recent results that have led to substantially reduced uncertainties on the EBL intensity over a wide range of wavelengths from 0.1 \mic\ to 15 \mic.  
 }

\section{Introduction}

The opacity of intergalactic space to  \gray s due to EBL absorption encodes important information for a host of astrophysical topics and also allows for tests of fundamental particle interactions.  From the perspective of \gray\ astronomers, detailed knowledge of the EBL is critical for our understanding of relativistic jets in distant \gray\ blazars and Gamma Ray Bursts, since it is required for the opacity correction~\cite{gs1967} of the observed  \gray\ spectra, and thereby  unveils the intrinsic spectra of these enigmatic sources.   

\begin{figure}
\begin{minipage}{0.99\linewidth}
\centerline{\includegraphics[width=0.9559995\linewidth,draft=false]{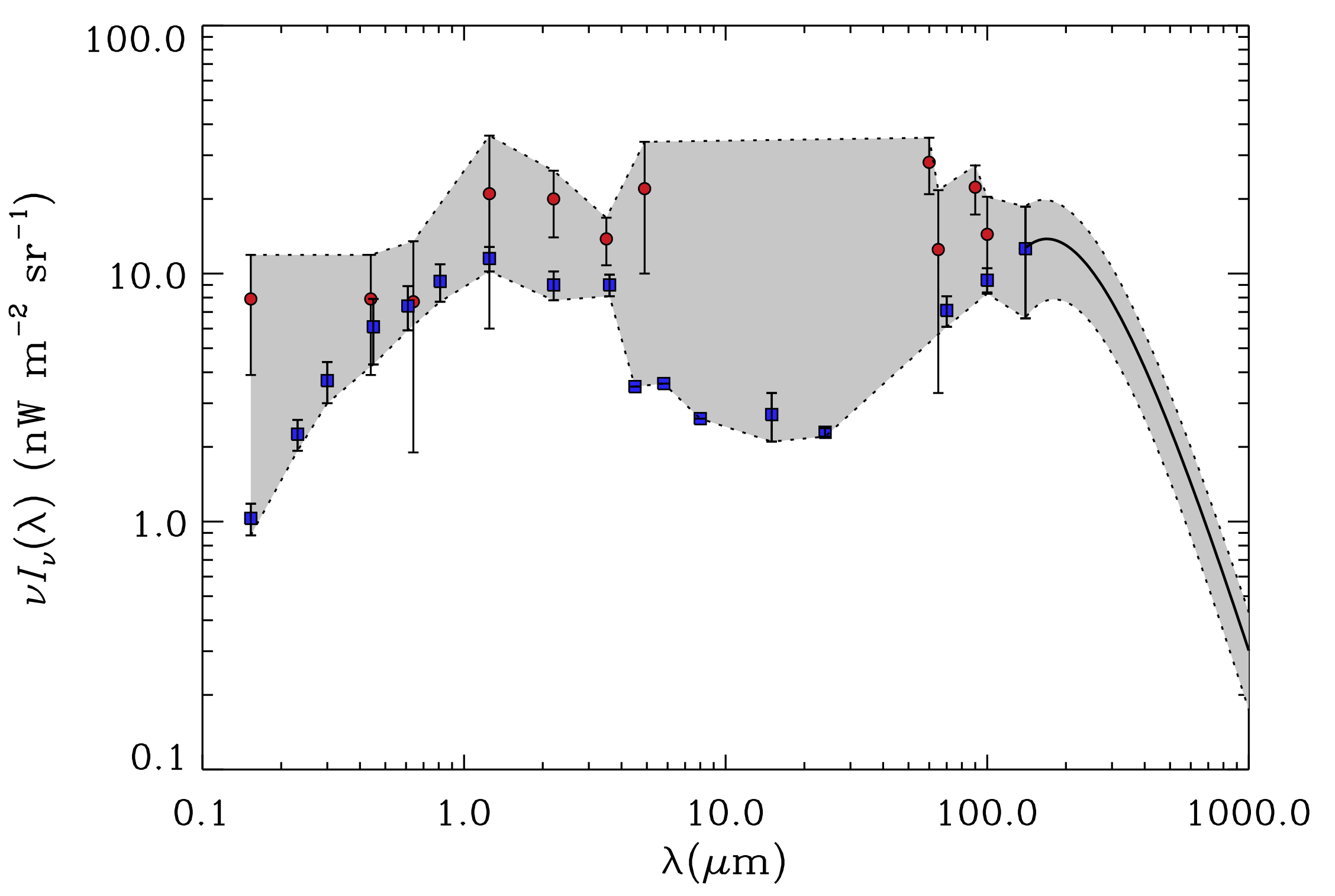}}
\end{minipage}
\caption[]{A range of EBL intensities from the UV to sub-millimeter wavelengths is shown (from Dwek \& Krennrich 2013 and references therein). 
The blue squares indicate lower limits from galaxy counts.  The red circles show absolute measurements.  The shaded area 
indicates the  EBL intensity allowed by a wide range of  UV to sub-millimeter observations.  }
\label{fig:EBLlimits}
\end{figure}

In a broader astrophysical context,  the EBL is a depository of all radiative energy releases since the time of decoupling, and is the second-most dominant diffuse radiation component that permeates our universe, right after the cosmic microwave background.  With star formation and accretion in active galactic nuclei (AGN) providing known contributions to the EBL,  it also plays an important role in cosmic consistency tests, e.g., by comparing it with related diffuse radiation fields~\cite{dk2013} including the X-ray background (AGN activity), the radio background and the cosmic supernova neutrino background (star formation).   Its spectrum is bimodal (see Fig.~\ref{fig:EBLlimits}) with one component peaking at $\sim 1$~\mic\ and comprising energy releases associated with the formation of heavy elements and the accretion of matter onto black holes in AGN. A second component peaking at $\sim 100$~\mic\ consists of absorbed UV and optical radiation that is re-radiated by dust at infrared (IR) wavelengths. The peaks are separated by a trough around $\sim 15$~\mic\  caused by the decrease of the stellar emission towards mid-IR wavelengths, and the rise in the dust emission spectrum towards far-IR wavelengths.

Absolute measurements of the EBL intensity continue to be complicated by the difficulties of subtracting the bright foreground radiation from zodiacal light and diffuse light from our galaxy~\cite{hd2001}.  A summary of absolute measurements is depicted in Fig.~\ref{fig:EBLlimits} as red circles with large uncertainties.   Strict lower limits to the EBL are given by galaxy counts and constrain the minimal EBL intensity (blue squares in Fig.~\ref{fig:EBLlimits}).    As a result, the shaded area of possible EBL intensities in Fig.~\ref{fig:EBLlimits} is well constrained from below, but is very broad owing primarily to the difficulties in performing absolute measurements in the UV/optical/IR  regime, and particularly in the mid-IR.   

Independently, the large range of possible EBL intensities has been  narrowed down substantially through recent analyses of  the \gray\  spectra of blazars.    Upper limits to the EBL arise from the opacity of the Universe to  \gray s  of  a wide range of energies.  While they are somewhat dependent on assumptions about the  intrinsic source spectra,  the substantial increase of the catalog of extragalactic \gray\  sources combined with a range of novel analysis methods~\cite{dk2005,aha2006,mr2007,okd2011,ack2012,abr2013} have yielded strong constraints to the EBL intensity.

Besides improving our knowledge about the EBL, discerning the \gray\ opacity of the universe offers a unique new opportunity for astroparticle physics at the intersection between non-thermal particle phenomena and thermal radiation fields. Anomalous features of the  \gray\ opacity have  the potential to reveal physical processes  that go beyond the standard models of particle physics and/or astrophysics.   For example, a significant discrepancy between EBL lower limits from galaxy counts and \gray\ opacity constraints (upper limits),  could provide hints of physics beyond the current realm of particle physics and astrophysics.  Such hints might include  putative radiation components \cite{ed1998}  due to primordial particle decay, or associated with Pop-III stars~\cite{dka2005} or dark stars~\cite{mau2012}, which are not accounted for in current galaxy counts,  and could persist as a residual background that increases the \gray\ opacity,  thus preventing upper limits from \gray\ data and galaxy counts from converging.   

Moreover, if the EBL derived from \gray\ opacity measurements were to fall consistently below the lower limits from galaxy counts would be a tantalizing result, which could be explained either by interactions of photons with axion-like particles (ALPs)~\cite{hs2007} or might result from secondary \gray\ rays produced in cosmic-ray cascades~\cite{ek2010} associated with the primary sources.  While the possibility of  detecting an ALP signature in   \gray\ opacity measurements contributes to dark matter searches, it  is hypothetical at this stage (but see also ~\cite{hm2012}).  The parameter space covered by these measurements  includes the mass between $\rm 10^{-12} - 10^{-7} eV$ and  probes coupling constants between $\rm 10^{-13} - 10^{-10} \: GeV^{-1}$, which is complementary to other  ALP searches.     Similarly, the detection of secondary \gray\ rays from ultra-high-energy cosmic-ray cascades from blazars implies two corollaries since their relativistic jets have  to accelerate hadrons to 10s of PeV and  secondary \gray\ rays from these cascades  are only detectable if the intergalactic magnetic fields are very small, of order $\rm 10^{-15}$~G or less.   

Due to the hypothetical nature of these processes we assume in the following that pair production from  \gag\ interactions is the only process relevant for the \gray\ opacity of our Universe.  This approach of using the minimal assumptions about the astrophysical contributors to the EBL and established physical processes allows one  to perform cosmic consistency tests and in the genuine absence of any new physics alleviating the EBL opacity,  also results in reliable EBL constraints.

\section{Constraints of the Gamma-ray Opacity }
The cross section for the $\gamma-\gamma$ interactions (see e.g., Dwek \& Krennrich~\cite{dk2013})   is broad compared to the energy resolution (15 - 20\%)  of space-based and ground-based \gray\  telescopes. The  cross section peaks at energies $E_{\gamma}$(TeV)$\approx 0.8\, \lambda$(\mic), so $\sim 1$~TeV photons are effectively attenuated by $\sim 1$~\mic\ photons from the EBL. 

\begin{figure}
\begin{minipage}{0.5\linewidth}
\centerline{\includegraphics[angle=-90, width=1.050099994\linewidth]{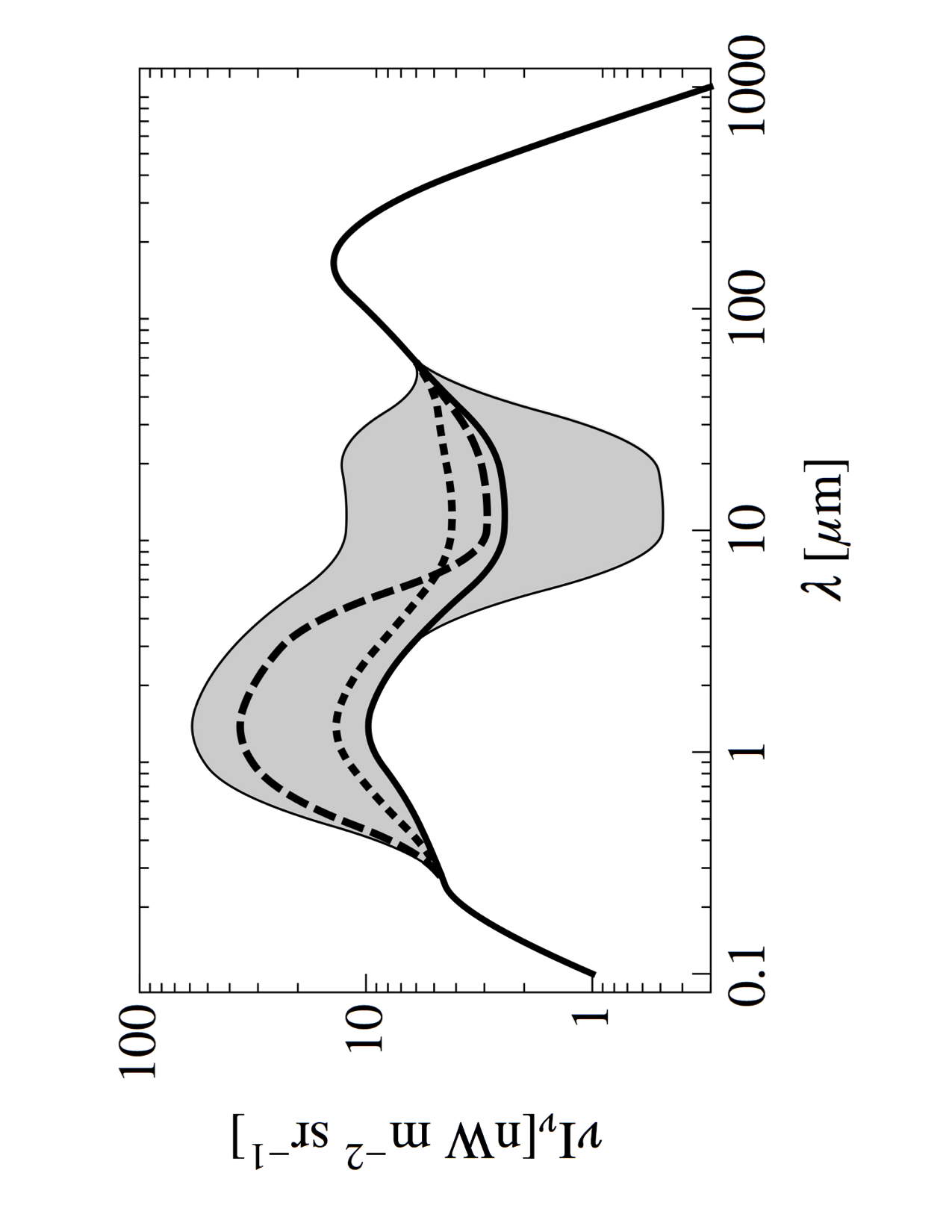}}
\end{minipage}
\hfill
\begin{minipage}{0.5\linewidth}
\centerline{\includegraphics[width=1.05000999395\linewidth,draft=false]{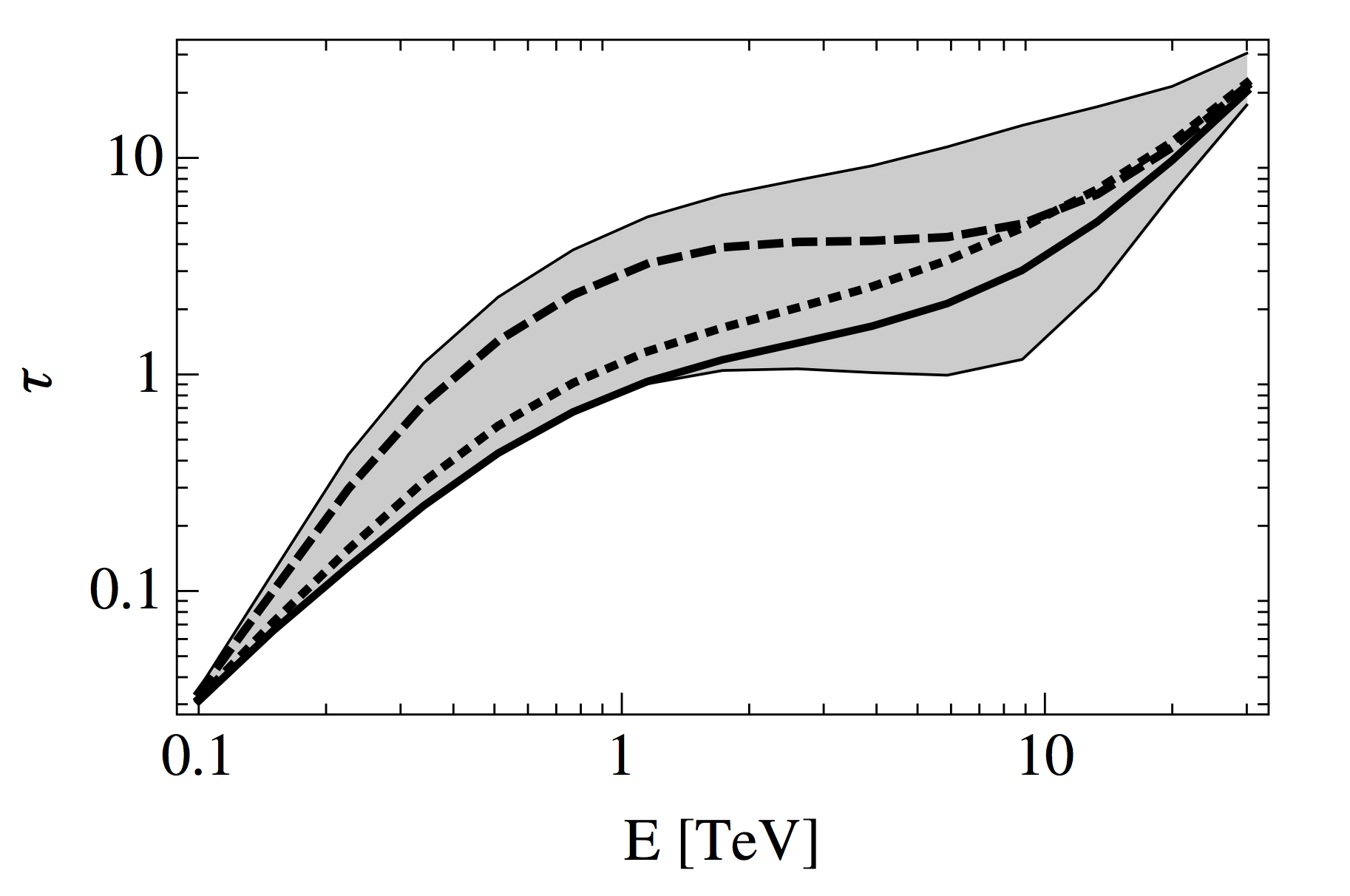}}
\end{minipage}
\caption[]{Left: EBL intensity vs. photon wavelength. The shaded region indicates the range of EBL scenarios. 
The thick solid line indicates  a baseline shape with a minimal near-IR intensity, For clarity, two additional models are shown (dotted and dashed) illustrating the independent scaling of the near- and mid-IR regions. Right: optical depth $\rm \tau$ (at z = 0.1) vs. gamma-ray energy in TeV for each EBL scenario are shown.  The optical depths for the baseline compared to the dashed line indicates a large near-IR producing a large rise in the opacity below 1 TeV, whereas a large near-IR to mid-IR ratio leads to a large change in the slope of  $\rm \tau$  around 1 TeV.  Figures are from Orr et al. 2011~\cite{okd2011} .}
\label{fig:EBL-Breaks}
\end{figure}

Multiple spectral imprints from absorption by the EBL are expected to occur between  10~GeV and  50~TeV.   The magnitude of the \gray\ opacity depends on the EBL intensity, and its energy dependence is determined by the spectral shape of the EBL. For an EBL intensity, $I_{\nu}$ that is given by a power law, e.g., $\nu I_{\nu}(\lambda) \sim \lambda^{\alpha}$, the energy dependence of the \gray\ optical depth is $\tau_{\gamma\gamma}(E_{\gamma})=E_{\gamma}^{\alpha+1}$.  Therefore, changes in the slope of the EBL intensity with wavelength will give rise to changes in the slope of the \gray\ opacity with energy $E_{\gamma}$ (see Fig.~\ref{fig:EBL-Breaks}, right).   

For example, the rise in the EBL intensity between the UV/optical (0.1 -- 0.5 \mic) to  near-IR  ($\rm \propto $~1 \mic) amounts to a redshift dependent  absorption feature detectable in  \gray\  spectra between 10 GeV to several 100 GeV, resulting in a gradually more prominent spectral break for higher redshift sources.    A second and more subtle spectral break (softening or hardening)  in  \gray\ spectra is expected at $\sim 1$~TeV.     This feature arises from a substantial drop in the EBL photon number density between the stellar/AGN emission component  at $\sim 1$~\mic\  towards the mid-IR  ($\sim 10$~\mic);  the corresponding  change in the slope (hardening in this case) of the \gray\ optical depth occurs around $\sim 1$~TeV (dashed line in Fig.~\ref{fig:EBL-Breaks}, right).  A third  spectral break is expected from the intensity  rise between the mid-IR trough and the far-IR EBL and the associated rise in the opacity.  The result would be a  spectral  softening in the 
    10 - 50 TeV energy regime.

\section{Blazars for Searching for EBL Absorption }

\begin{figure}
\begin{minipage}{0.8\linewidth}
\centerline{\includegraphics[width=1.0090\linewidth,draft=false]{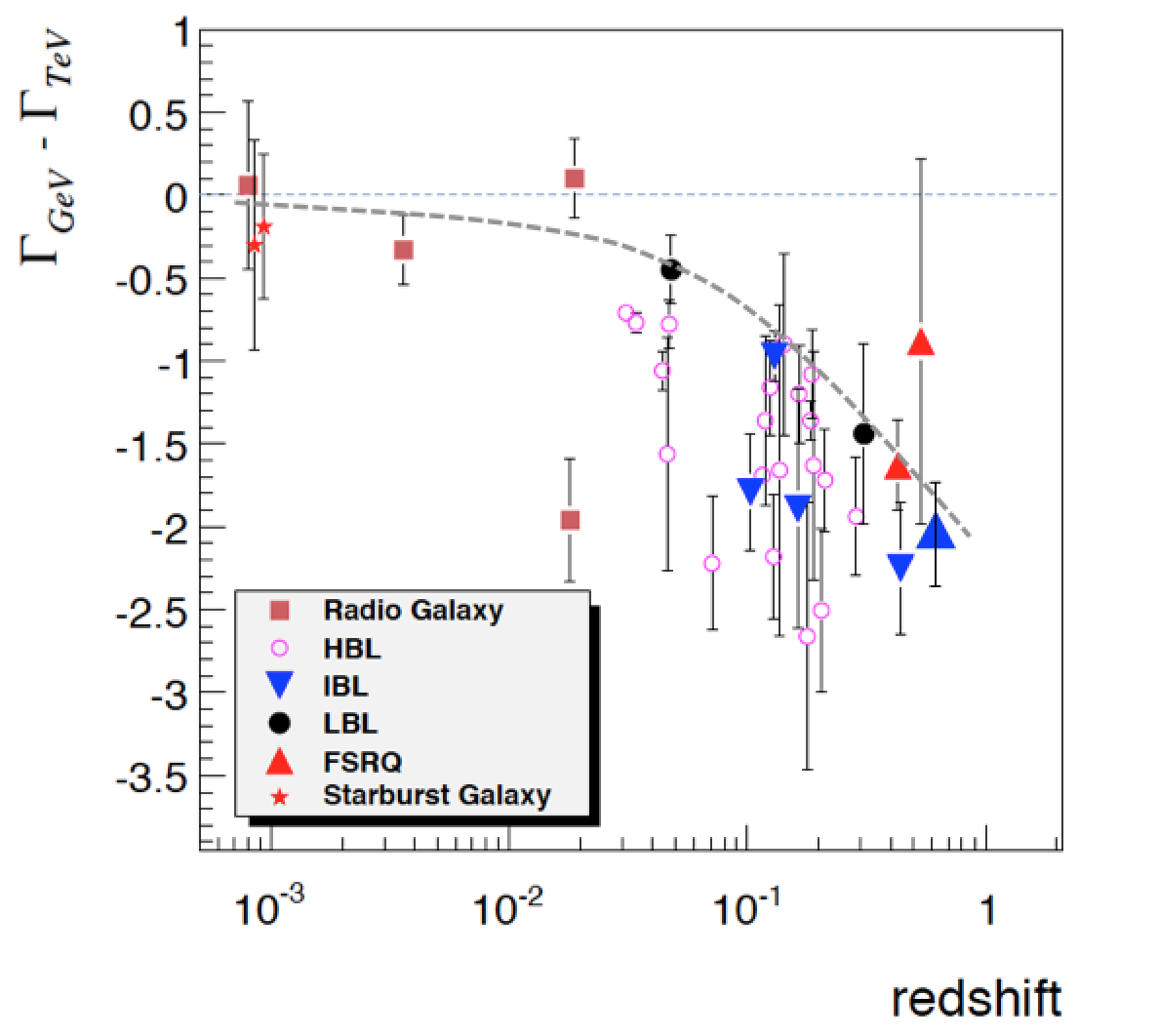}}
\end{minipage}
\caption[]{The difference between $\rm \Gamma_{GeV}$, the spectral index at GeV (\fermilat) energies, and  $\rm \Gamma_{TeV}$, the energy spectral index in the TeV regime (H.E.S.S, MAGIC, VERITAS) is shown as a function of their redshift. Red squares (radio galaxies), red stars (starburst galaxies), empty circles (HBLs, high-frequency peaked BL Lacs), blue downward triangles (intermediate-frequency peaked BL Lacs), filled circles (LBLs, low-frequency peaked BL Lacs), red upward triangles (FSRQs, flat spectrum radio quasars) indicate the different types of \gray\ sources.  The figure has been adapted from Dwek \& Krennrich 2013, however, a recent data point of PKS~1424+240 has been added (blue upward triangle), with the caveat that the redshift of the source is a lower limit.}
\label{fig:Blazar-Breaks}
\end{figure}

Blazars  currently provide the largest sample of extragalactic \gray\ sources  to search for spectral signatures from EBL absorption.  Collectively, the \fermilat\ and   imaging atmospheric Cherenkov telescopes (IACTs, such as \hess, \magic\ and \veritas)  provide  EBL-relevant energy coverage from 10s of GeV to 10s of TeV.   Fig.~\ref{fig:Blazar-Breaks}  shows  the change of the spectral slope between the \gray\  spectral index $\rm \Gamma_{GeV}$ in the GeV regime and the  spectral index $\rm \Gamma_{TeV}$ in the TeV regime for a subset of $\rm \sim$~3 dozen extragalactic sources~\cite{dk2013}. 

A clear trend showing spectral softening with increasing redshift  (z = 0.0008 to $\rm \sim$~0.6.) is apparent.  There is also a considerable variance in the magnitude of the spectral break for a given redshift.  This arises from spectral steeping in the GeV to TeV regime intrinsic to some of the sources.    Sources closest to the dashed line are the ones whose spectral break is dominated by EBL absorption, and exhibit little or no source intrinsic spectral steepening.  The latter include hard spectrum blazars such as  1ES~1101-232, 1ES~1218+304, 1ES~0229+200, RGB~J0710+59, already indicated by their unusually hard energy spectra~\cite{aha2006,dk2013} given their substantial redshift.      

 Typical broad-band blazar spectra  in the radio to the  \gray\ regime have  two  emission peaks in $\rm \nu F_{\nu}$, one in the radio-to-X-ray waveband and a high energy peak in the GeV to TeV  \gray\ regime.   In some cases, the blazar emission can be convincingly modeled as synchrotron-self-Compton (SSC) emission.  In the SSC model, the high energy emission is produced by inverse-Compton scattering of synchrotron photons emitted by a common population of electrons.  However ambient  soft photons can also  contribute to the target for inverse Compton scattering and complicate the modeling of the \gray\ peak considerably.  In addition, blazar variability combined with the difficulty of getting contemporaneous multi-wavelength coverage has prompted approaches that constrain the   \gray\ peak using general features.

\begin{figure}
\begin{minipage}{0.9\linewidth}
\centerline{\includegraphics[angle=-0, width=1.040099994\linewidth]{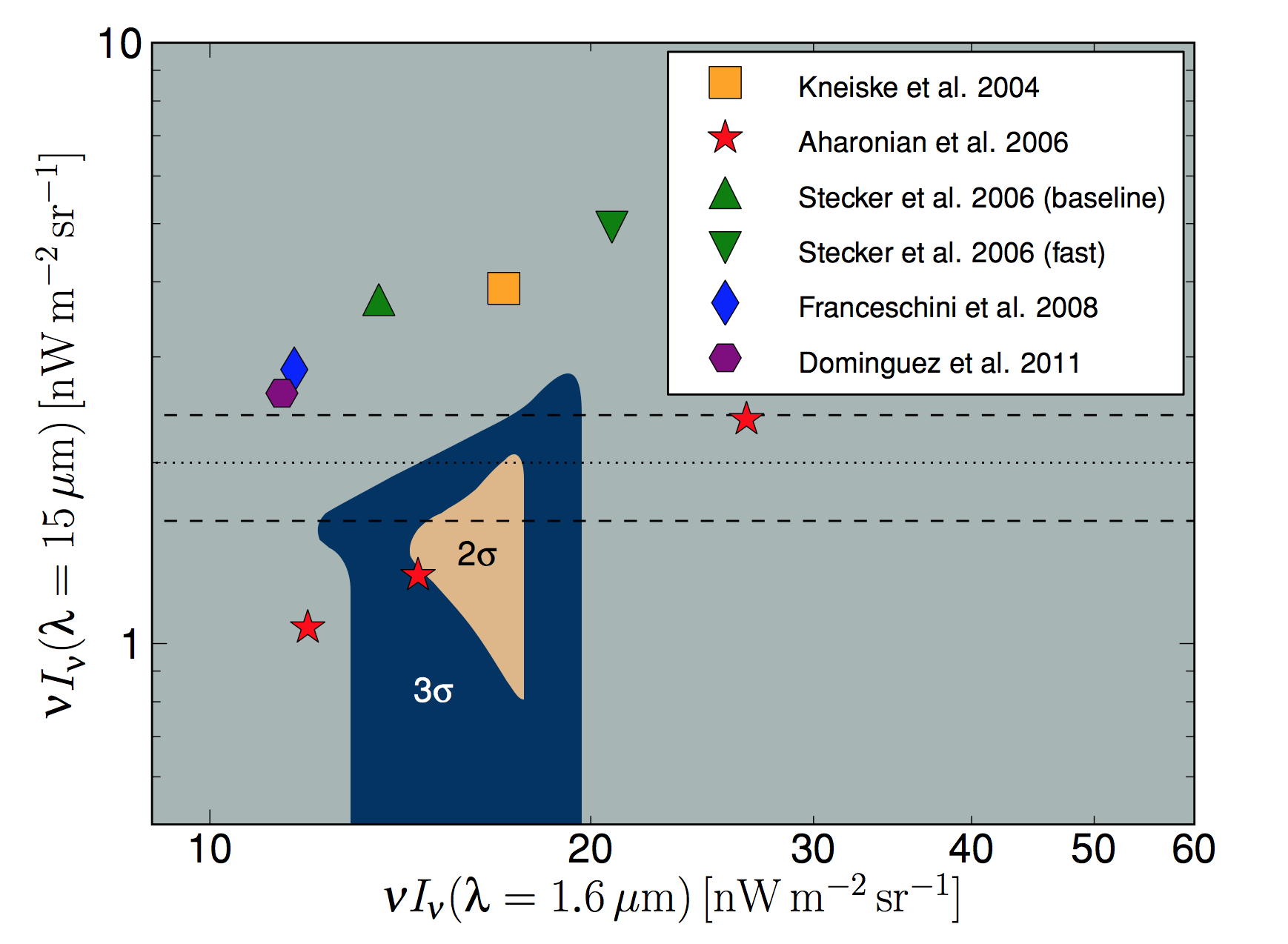}}
\end{minipage}
\caption[]{EBL constraints derived for an analysis~\cite{okd2011} based on two spectral breaks independently sensitive to the near-IR and mid-IR EBL are shown.  The combined 2$\rm \sigma$ and 3$\rm \sigma$ contours from this analysis limit the near-IR to mid-IR ratio substantially.  Also shown for reference are several EBL  model predictions.}
\label{fig:EBL-Orr}
\end{figure}

Over a small energy regime ($\rm \sim$ magnitude in energy), the  \gray\ spectra are generally well approximated  by power laws.  On a larger energy scale they exhibit a concave shape (e.g. parabolic, broken power law, exponential cutoff) and it is this empirical feature that can be used  to constrain the intrinsic spectra. The relation between the intrinsic,  and the observed  blazar spectrum is given by:
   $\rm (dN/dE)_{int}   =   (dN/dE)_{obs}  \: \:   e^{ \tau_{\gamma \gamma} (E, z)}$.
The equation implies that an overestimate of the  opacity will lead to an exponential rise in the inferred intrinsic spectrum of the blazar.  Such exponential rise is unphysical. It is inconsistent  with our basic understanding of blazars, and is absent in the observed  \gray\ spectra of blazars for which $ \rm \tau_{\gamma \gamma} (E, z)$ is negligible.    Similarly, a theoretical argument made by Aharonian et al. (2006)~\cite{aha2006} that is based on theory of diffusive shock acceleration, suggests that the intrinsic energy spectra in the gray\  regime cannot be harder than $\rm \Gamma_{TeV} \ge 1.5$.

\section{Recent EBL Constraints from Opacity Measurements -- Search for Unique Spectral Signatures  }

\begin{figure}
\begin{minipage}{0.95\linewidth}
\centerline{\includegraphics[angle=-0, width=1.13703399994\linewidth]{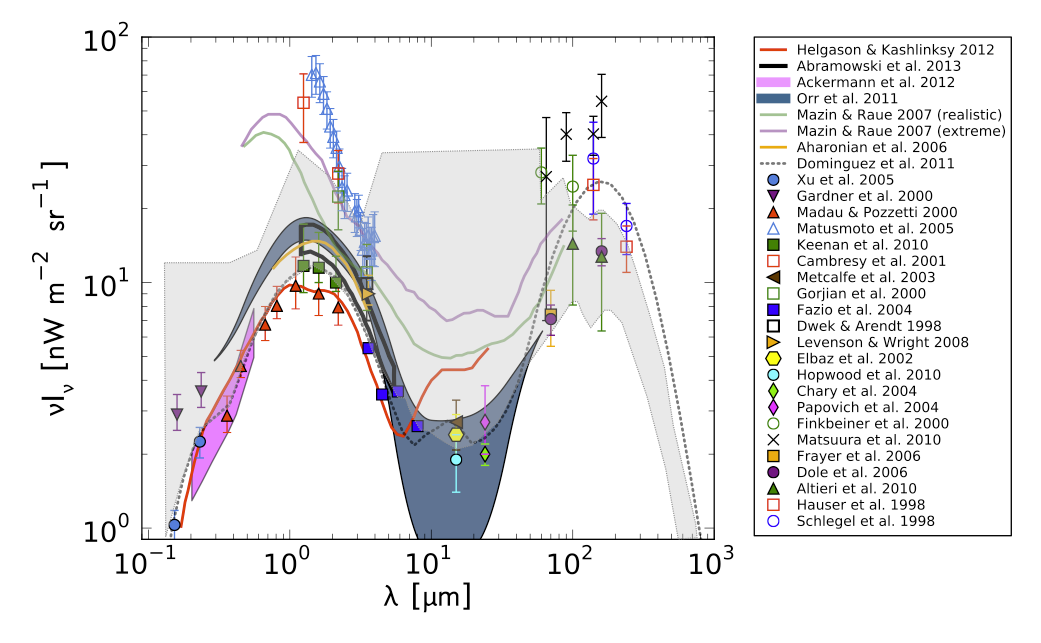}}
\end{minipage}
\caption[]{Summary of the EBL from direct measurements (open symbols), lower limits from galaxy counts (filled symbols), observed galaxy luminosity functions  (legend listing 1), constraints from IACT observations of blazars (legend listings 2-7), and the model of Dominguez et al. (2011)\cite{d2011} (legend listing 8). Other references can be found in Dwek \& Krennrich (2013)\cite{okd2011}. The possible range of the  EBL intensity (gray shaded) adapted from Fig.~\ref{fig:EBLlimits}  has been largely reduced.}
\label{fig:EBL-Summary}
\end{figure}

Several different methods for searching for evidence of EBL absorption in blazar spectra have been performed. The most recent ones, providing the strongest constraints, include results by the  \fermilat\ collaboration~\cite{ack2012}, and the \hess\  collaboration~\cite{abr2013} and results by Orr et. al.~\cite{okd2011}.       These results use different techniques.  The \fermilat\ result uses $\rm \sim$ 150 BL Lacs (sub class of blazars) spanning a redshift range of z = 0.03 - 1.6 and a global fit function  for the observed spectra that are modified by $\rm  \propto e^{-b \: \tau_{model}} $,  with b being a free parameter  that is constrained by the data to b=1 ($\tau(E, z) = b \:  \tau_{model} $), showing that EBL absorption is taking place.  This result  is most constraining for  the EBL at  0.3 \mic .  The energy dependent cutoff feature observed in the energy spectra of this large blazar sample can be  regarded as a detection of an EBL signature in the 10~GeV - 100s GeV regime associated with the strong rise of the EBL intensity in the UV/optical toward larger wavelengths.  These constraints are indicated in Fig.~\ref{fig:EBL-Summary} as a shaded (magenta) band at short wavelengths.

IACT results are based on substantially smaller samples.  The work by Orr et al. (2011)~\cite{okd2011} uses a sample of 12 blazars between redshifts of 0.044 - 0.186 to constrain the near-IR,  and mid-IR EBL intensities and the near-IR to mid-IR ratio. The technique used consists of two parts. One is designed to constrain the second spectral break discussed in Section 2 of this paper by measuring the spectra below and above the expected break energy of $\rm \sim$~1 TeV, thereby constraining the near-IR to mid-IR ratio.  Furthermore, by using the combination  of \fermilat\  and IACT data for 4 hard-spectrum blazars, the near-IR intensity is strongly constrained under the assumption that the energy cutoffs in these spectra are dominated by EBL absorption, which is well justified given their position in Fig.~\ref{fig:EBL-Breaks} close to the dashed gray line.  This work led to a well-constrained region of the near-IR to mid-IR ratio shown in 
Fig.~\ref{fig:EBL-Orr}.   These results also provide strong constraints to the absolute  near-IR and mid-IR  intensities and the possible range of EBL scenarios consistent with these data are  also shown   in Fig.~\ref{fig:EBL-Summary} as a shaded blue regime.

Results presented by the \hess\ collaboration~\cite{abr2013}  are based on a technique similar to the one applied by the \fermilat\  collaboration using a variable that allows the EBL attenuation term to scale  $\rm  \propto e^{-b \: \tau_{model}}$, and provide a significant detection of an EBL absorption feature.   These results are also shown in Fig.~\ref{fig:EBL-Summary}  as a region bounded by the black solid line. 

As can be seen, opacity measurements with \gray\ telescopes have provided strong constraints on the EBL and have helped to substantially reduce the uncertainties affecting absolute EBL measurements.      While the UV EBL discerned from \gray\ data is consistent with lower limits from galaxy counts, the near-IR EBL leaves some room for additional EBL contributions not accounted for in galaxy surveys.  

Further \gray\ studies with substantially larger blazar samples and fully resolved galaxy counts are required to reduce the statistical and systematic uncertainties of both approaches.   The next generation Cherenkov Telescope Array (CTA)\cite{actis2011}  and the James Webb Space telescope (JWST)\cite{JG2006} will be required to achieve convergence between lower limits and opacity measurements.    Large samples of blazars also have the potential to provide much-improved constraints on the EBL in the optical/near-IR and mid-IR through a better understanding of the blazar subclasses  and their intrinsic spectra, as well as better photon statistics for the measurement of the redshift dependence of any spectral feature attributable to the EBL.   The JWST will likely resolve the EBL sources at near-IR wavelengths due to its unprecedented resolution and sensitivity.

\section*{Acknowledgments}

This research is supported by grants from the U.S. Department of Energy Office of Science, and by NASA Fermi Guest Investigator grant NNX11AO38G 
and by Iowa State University.   Furthermore, I would like to thank Hugh Dickinson for proof-reading the manuscript.

\end{document}